\documentclass[12pt,a4paper]{iopart}
\usepackage{graphicx}

\begin{document}

\title{Defining and controlling double quantum dots in single-walled carbon nanotubes}
\author{MR Gr{\"a}ber, M Weiss, and C Sch{\"o}nenberger}
\address{Institut f{\"u}r Physik, Universit{\"a}t Basel,
Klingelbergstrasse 82, CH-4056 Basel, Switzerland}
\ead{christian.schoenenberger@unibas.ch}

%\pacs{73.63.-b, 73.23.-b, 03.67.-a} \keywords{carbon
%nanotube,double quantum dot, molecular electronics, quantum
%computing, local gate control}

\begin{abstract}
We report the experimental realization of double quantum dots in
single-walled carbon nanotubes. The device consists of a nanotube
with source and drain contact, and three additional top-gate
electrodes in between. We show that, by energizing these
top-gates, it is possible to locally gate a nanotube, to create a
barrier, or to tune the chemical potential of a part of the
nanotube. At low temperatures we find (for three different
devices) that in certain ranges of top-gate voltages our device
acts as a double quantum dot, evidenced by the typical honeycomb
charge stability pattern.

\end{abstract}

% \submitto{SST} 

\maketitle

\section{Introduction}

Since their discovery in 1991 \cite{Iijima1} carbon nanotubes
have, due to their unique mechanical and electronic properties,
been subject of a tremendous scientific and technological
interest. In the field of mesoscopic physics, carbon nanotubes
offer an easily accessible experimental platform for studying the
physics of the text book example of a particle trapped inside a
box, a so-called quantum dot or artificial
atom~\cite{kastner,bockrath}. Single quantum dots can simply be
realized by contacting a nanotube with two metallic contacts
(normally made of Palladium); the contacts between the nanotube
and the metallic leads usually act as tunnel barriers,
characterized by the nanotube-lead tunneling rate $\Gamma$ and a
capacitance $C$. The energy scales for nanotube quantum dots are
given by a typical single-electron charging energy $U_{C}\approx
3$~meV~$\approx 30$~K and a quantum-mechanical level spacing
$\delta E=\frac{h v_F}{2L}$, where $h$ is Planck's constant. Using
the Fermi velocity $v_{F}=8 \times 10^{5}$~m/s and an effective
nanotube length $L=1\,\mu$m the level spacing amounts to $\delta E
\approx 2$~meV.

However, the standard approach for manufacturing quantum dot
devices has relied on structures in GaAs-based 2-dimensional
electron gases (2-DEG), which can be defined using etching and
gating techniques. The main advantage of this system is the high
degree of control over the quantum dot properties, which has been
achieved over the last years. These quantum dots allow for a
precise tuning of the coupling to the leads by energizing locally
acting gate electrodes, see e.g. \cite{KouwenhovenReview} and
references therein. Additionally, center gates can be used in
order to define double quantum dot structures with a tunable
inter-dot coupling. This tunability is an essential ingredient for
further experiments exploring the nature of electronic states in
quantum dots - or, even more ambitious, for realizing quantum
electronic devices such as spin- or charge-based quantum
bits~\cite{loss,schoen,Burkard}. Whereas this high degree of
control has been lacking in nanotube-based quantum dots so far,
using carbon nanotubes offers fascinating opportunities. For
example, new physical phenomena such as superconducting
correlations or spin injection into quantum dots can be studied in
carbon nanotube quantum dots~\cite{buit2,Sahoo}. In contrast to
carbon nanotubes, up to now it has not been possible to attach
ferromagnetic and superconducting to GaAs-based quantum dots.
Moreover, the influence of the surrounding nuclear spins is
expected to limit electron spin dephasing times in GaAs (double)
quantum dots~\cite{hyperfine}. In carbon nanotubes, on the other
hand, nuclear spins are predominantly absent and hyperfine
interactions thus strongly reduced. The question, to which degree
carbon nanotube quantum dots can be tuned using locally acting
gate electrodes is therefore an important issue to address. In
this article we describe a technique of implementing local
top-gate electrodes onto a single-walled carbon nanotube (SWNT).
After characterizing the functionality of the top-gates we will
then make use of them in order to define and control double
quantum dots inside SWNTs.

%%%%%%%%%%%%%%%%%%%%%%%%%%%%%%%%%%%%%%%%%%%%%%%%%%%%%%%%%%%%%%%%%%%%%
%%%%%%%%%%%%%%%%%%%%%%%%%%%%%%%%%%%%%%%%%%%%%%%%%%%%%%%%%%%%%%%%%%%%
%%%%%%%%%%%%%%%% Section local gates %%%%%%%%%%%%%%%%%%%%%%%%%%%%%%%%
%%%%%%%%%%%%%%%%%%%%%%%%%%%%%%%%%%%%%%%%%%%%%%%%%%%%%%%%%%%%%%%%%%%%%
%%%%%%%%%%%%%%%%%%%%%%%%%%%%%%%%%%%%%%%%%%%%%%%%%%%%%%%%%%%%%%%%%%%%%

\section{Local gating of carbon nanotubes}

\subsection{Strategies for gating nanotube quantum dots}

With the nanotube lying on an oxidized Si-substrate, a natural way
of gating this single quantum dot is to apply a voltage to the
doped Si-substrate. The Si then acts as a back-gate globally
affecting the whole quantum dot. In order to create multiple dots
in such a device and control them independently, however, one will
need to find a way of locally gating a nanotube. By using such
local gates, one can either create a barrier, or simply shift the
chemical potential within a small part of the nanotube. In the
following, we will briefly review two different strategies of
local gating of nanotubes that have been reported in the
literature, and will describe in detail the technique that has
been developed in our lab. At the end of this section measurements
of electrical transport through nanotube devices with local gates
will be presented.

\begin{figure}
\begin{center}
\includegraphics*[width=0.8\linewidth]{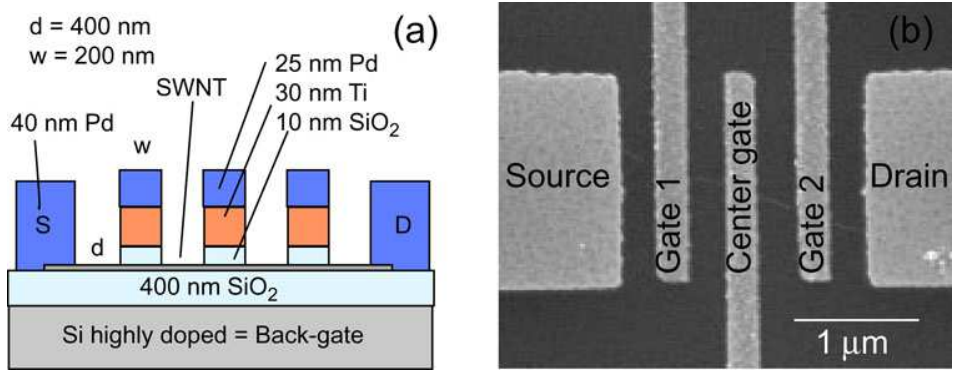}
\caption{\label{figure1} (a) Side-view schematic of a SWNT device
with three top-gates. (b) Scanning electron micrograph of the
device. Gates are labelled gate~1, center~gate, and gate~2 (from
source to drain).}
\end{center}
\end{figure}

In order to fabricate gate electrodes locally acting on a
nanotube, side-gates represent a straightforward
option~\cite{BiercukPRL}. Besides source and drain contact,
additional electrodes are patterned in the vicinity of the
nanotube. The advantage of this technique is that contacts and
side-gates can be fabricated within the same processing step. It
is, however, difficult to get the side-gates as close to the
nanotube as possible, yet not contacting it electrically. Thus,
typically side-gates are spaced by approximately 100~nm from the
SWNT, making the gating less efficient and their action less
local.

More efficient are gates made by directly evaporating the gate
electrode on top of the nanotube, with a thin gate oxide
underneath. These so-called topgates are spaced from the SWNT only
by the thickness of the gate oxide ($\approx 1-10~nm$), making
them act more efficiently and (depending on their width) more
locally as compared to side-gates. Despite the fact that there are
drawbacks of this method as well (additional processing steps,
nanotube properties may be modified underneath the top-gates),
top-gates are the most promising approach for creating local
barriers in SWNT. Therefore, we have developed a reliable method
for fabricating top-gate electrodes in our laboratory, which we
will now discuss in more detail.

\subsection{Experimental}

SWNTs were grown on a degenerately doped Si/SiO$_2$ substrate by
means of chemical vapor deposition (CVD). Details of the CVD
process can be found elsewhere~\cite{Juerg}. After the initial
preparation of SiO$_2$/Ti/Au bond pads and alignment markers,
SWNTs were then localized with a scanning electron
microscope~(SEM). In the following step the gate electrodes were
defined by e-beam-lithography. Electron-gun-evaporation of SiO$_2$
as gate-oxide, Ti as gate-metal, and Pd serving as anti-oxidant
cover layer followed. The gate-oxide film thickness was chosen to
be 10~nm, the Ti film thickness 30~nm, and that of the Pd layer
25~nm. The materials were evaporated at a pressure of
$\approx$~10$^{-7}$~mbar. In a final lithography and evaporation
step the source and drain electrodes of the nanotube, consisting
of 40~nm Pd, were defined. The evaporation conditions were the
same as described above, except the substrate was kept at a
constant temperature of $\approx$ 0$^{\circ}$ C by cooling the
sample holder inside the evaporation chamber. This cooling helps
to reduce outgasing of materials inside the vacuum chamber due to
heating during the evaporation. After lift-off of the remaining
PMMA, the samples were glued into a 20-lead chip carrier and
bonded. Figures~\ref{figure1}(a) and (b) show a side-view
schematic and a scanning electron micrograph of a typical SWNT
device with three top-gates in addition to the source and drain
electrode. The spacing between source and drain electrode amounts
to 2.2~$\mu$m, and the width of the gates was chosen to be 200 nm.
The back-gate oxide has a commonly used thickness of 400~nm.

\begin{figure}[t]
\begin{center}
\includegraphics*[width=\linewidth]{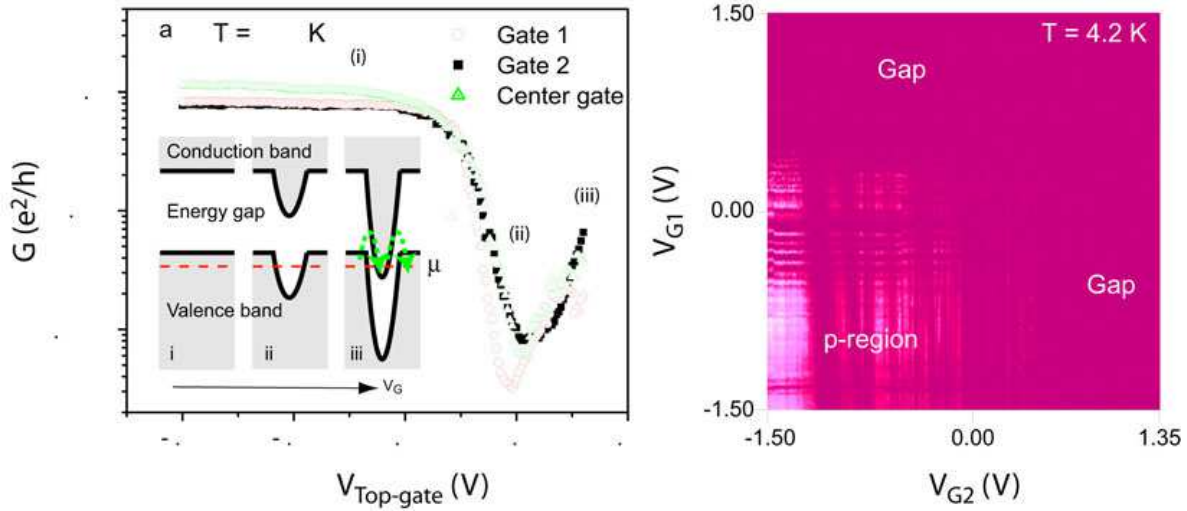}
 \caption[Bending of bands due to local gates]{(a) Linear conductance G on a logarithmic scale for a
  device with three top-gate electrodes (oxide thickness 10 nm) versus top-gate-voltage at T = 300~K.
 The gates non swept are connected to ground potential. Inset: (i) - (iii) illustrate the band structure for
 increasing top-gate voltage. (b) Colorscale plot (dark=0, bright=0.008~e$^2$/h) of the conductance versus
 gate~1 and gate~2 for constant center-gate voltage at 4.2 K.} \label{figure2}
\end{center}
\end{figure}

\subsection{Effect of local gate electrodes at 300~K and at 4.2~K}

In Fig.~\ref{figure2}(a) the linear conductance versus gate
voltage of a device with three top-gate electrodes is plotted. The
gate-dependence identifies the semiconducting nature of the SWNT.
At a voltage of roughly 0.6 V applied to either of the three
top-gates the conductance through the device is suppressed
indicating that the chemical potential is shifted locally into the
semiconducting gap of the SWNT. After a decrease of conductance
for increasing gate voltage, the conductance rises again for more
positive gate voltages. This behavior is explained by the band
diagram sketched in the inset of Fig.~\ref{figure2}(a).
Intrinsically the tube is p-doped and the chemical potential $\mu$
resides in the valence band (i). For increasing voltage at the
top-gate the potential landscape is changed locally, making $\mu$
lie within the energy gap below the gate (ii). In this scenario
the conductance through the nanotube reaches its minimum. With
this technique it should thus be possible to create local barriers
inside a carbon nanotube, allowing one to create artificial
potential landscapes. If the gate voltage is increased even more,
the lower edge of the conduction band will eventually reach the
upper edge of the valence band (iii). Now thermally activated
band-to-band processes indicated by the green arrows are possible
and the conductance increases again. We have observed such
behavior only at 300~K indicating the large activation barriers
involved in these band-to-band charge transfer processes.
Band-to-band charge transfer processes have also been reported in
Ref.~\cite{Appenzeller}.

In Fig.~\ref{figure2}(b) the linear differential conductance at
4.2~K is plotted on a colorscale (bright=more conductive) versus
voltages applied at the top-gates~1~and~2 for a constant
center-gate voltage of $V_{C}=-1$~V. At voltages of around 0.5~V
applied to either of the top-gates the chemical potential is
shifted into the energy gap of the nanotube and electrical
transport is suppressed. For lower top-gate voltages, sweeping
gate~1 and gate~2 leads to pronounced oscillations of the
conductance due to single-electron charging and finite-size
effects of the nanotube, which are accessible at low temperatures.

%%%%%%%%%%%%%%%%%%%%%%%%%%%%%%%%%%%%%%%%%%%%%%%%%%%%%%%%%%%%%%%
%%%%%%%%%%%% Double quantum dots %%%%%%%%%%%%%%%%%%%%%%%%%%%%%%
%%%%%%%%%%%%%%%%%%%%%%%%%%%%%%%%%%%%%%%%%%%%%%%%%%%%%%%%%%%%%%%

\section{Nanotube double quantum dots}

\subsection{Previous work}

Recently, in the field of double quantum dots in carbon nanotubes
an enormous progress has been achieved. In 2004 Mason et al. first
demonstrated the local gate control of an intrinsic double quantum
dot inside a carbon nanotube \cite{Mason}. This work was then
extended by the same group in Ref.~\cite{Biercuk}, where a tunable
mutual capacitance was demonstrated. In a recent work,
Sapmaz~et~al. could observe electronic transport through excited
states as seen in finite-bias triangles in a SWNT double
dot~\cite{sapmaz}. In Ref.~\cite{graeber} molecular eigenstates of
a strongly coupled carbon nanotube double quantum dot were
observed and analyzed.

\subsection{Experimental data}

\begin{figure}[t]
\begin{center}
\includegraphics*[width=\linewidth]{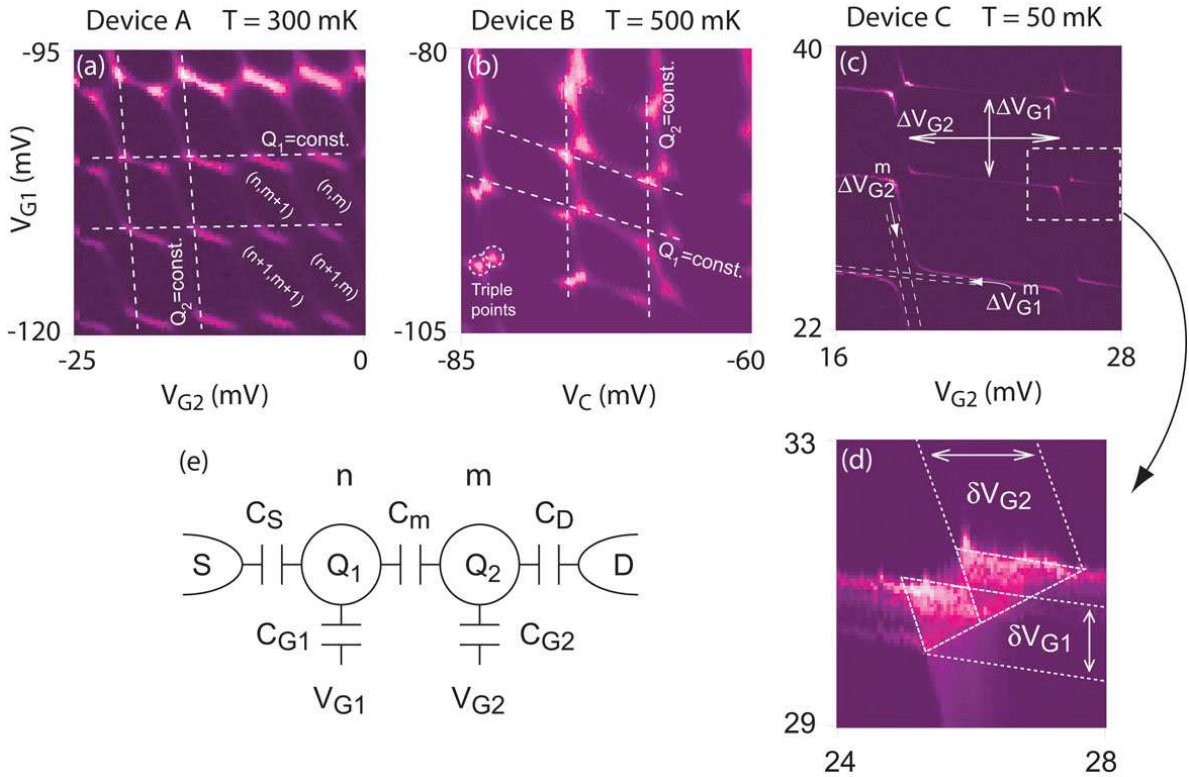}
\caption{(a) Colorscale plot of the conductance versus top-gate
voltages at 300~mK for device A. Bright corresponds to
0.4~e$^2$/h. The obtained honeycomb pattern is the charge
stability map of a double quantum dot. (b) Same for device B at
500 mK, bright corresponds to 0.08~e$^2$/h. (c) Same for device C
at 50 mK, bright corresponds to 0.035 e$^2$/h. (d) Zoom into the
triple point region marked by the dashed box in (c) at a bias
voltage of $V_{sd}$=500$\mu$V. (e) Capacitive model of a double
quantum dot.} \label{figure6}
\end{center}
\end{figure}

In this section we will show that it is possible to reliably
define clean double quantum dots in SWNTs by using top-gate
electrodes. We focus on three devices labelled A,B,C with three
top-gates each. Samples A and B were fabricated according to
Fig.~\ref{figure1}(a). In the case of device C the source -drain
spacing was reduced to 1.4~$\mu$m and the top-gate width to
100~nm. Whereas devices A and B are based on a semiconducting SWNT
(operated in the hole regime), device~C is metallic. In
Fig.~\ref{figure6}(a)-(c) the differential conductance versus
voltages applied at two top-gates is plotted on a colorscale
(bright=more conductive). For devices A and C the center gate has
been set to a constant value of -0.1~V and 0~V, respectively, and
gate~1 and gate~2 are swept. In case of device B, center gate and
gate~1 are swept, while gate~2 was kept at a constant voltage of
$V_{G2}=-0.1$~V. The visible high-conductance ridges as observed
for all three devices define a charge-stability map that is shaped
like a honeycomb. This honeycomb pattern is characteristic of a
double quantum dot. Within each cell, the number of holes (n,m) on
the two dots is constant. Energizing gate~1~(2) to more negative
voltages successively fills holes into dot~1~(2), whereas a more
positive voltage pushes holes out of the dot. The fact that all
three devices can be tuned to exhibit a honeycomb pattern shows
that the double quantum dots are indeed defined by the local gates
and not intrinsic to the nanotube. Common to all three devices is
that the applied gate voltages are close to 0~V, i.e. far off the
pinch-off voltage. In such a regime, we expect a smooth modulation
of the electronic potential rather than sharp and steep barriers.

The honeycomb charge stability map allows for a quantitative
determination of the double dot capacitances as defined in the
electrostatic double dot model in Fig.~\ref{figure6}(e), following
the work of van der Wiel et al.~\cite{VanderWiel}. As an example
we will determine the capacitances of the double dot defined in
device C, see Fig.~\ref{figure6}(c). From the dimensions of the
honeycomb cell one can extract the gate capacitances:
\begin{equation}
C_{G1/2}=\mid e \mid/\Delta\,V_{G1/2}\;\;\;,
\end{equation}
yielding $C_{G1}\approx30$~aF and $C_{G2}\approx25$~aF. Of
particular importance are the points where three charge states are
degenerate, so-called triple points. Two such points are marked by
dashed circles in Fig.~\ref{figure6}(b) for clarity. When applying
a finite bias voltage, the triple points transform into triangles,
in which transport is enabled. Fig.~\ref{figure6}(d) shows the
triple point region within the dashed box of Fig.~\ref{figure6}(c)
at an applied source-drain voltage of $V_{sd}=500\mu V$. From the
dimensions of these triangles $\delta V_{G1(G2)}$ and
\begin{equation}
C_{G1(G2)}/C_{1(2)}=\mid V_{sd}\mid/\delta V_{G1(G2)}\;\;\;,
\end{equation}
we obtain the total capacitance $C_{1}=C_s+C_{G1}+C_m\approx
60$~aF and $C_{2}=C_d+C_{G2}+C_m\approx 75$~aF of dot~1 and dot~2,
respectively. Here $C_m$ denotes the mutual capacitance and
$C_{s(d)}$ the capacitance of the tunnel barrier to source
(drain). In a purely electrostatic model the mutual capacitance
can be evaluated from the spacing of two adjacent triple points.
This spacing, however, is influenced by the tunnel coupling $t$ in
between the two dots as well. This quantum mechanical effect leads
to a level anti-crossing, resulting in curved wings in the
vicinity of the triple points. A rough estimate of the mutual
capacitance, however, can be achieved by drawing the asymptotes to
the curved borders of the honeycomb, see the bottom left triple
point region of Fig.~\ref{figure6}(c). From the vertical
(horizontal) distance $\Delta V_{G1}^{m}$ ($\Delta V_{G2}^{m}$) it
is then possible to extract $C_{m}$ by using
\begin{equation}
\Delta V_{G1,2}^{m} = |e| C_{m} / C_{G1,2} C_{2,1} \;\;\;.
\end{equation}
We obtain a mutual capacitance of $C_{m}\approx 5$~aF.
Additionally, analyzing the curvature of the honeycomb borders
allows one to precisely evaluate the tunnel coupling~$t$. For a
detailed description on this we refer to Ref.~\cite{graeber},
where it was found that $t$ can exceed the electrostatic
nearest-neighbor interaction by as much as an order of magnitude.
This fact reflects the one-dimensional geometry of a nanotube;
electrostatic interactions are reduced due to the large separation
of the `center of mass' of the charges (while still allowing a
significant overlap of the wavefunctions).

\subsection{Where exactly are the two dots?}

So far we have seen that it is possible to reliably define and
control double quantum dots in SWNTs - the question where
precisely the two dots are located, however, has not been
addressed yet. As we will point out, from Fig.~\ref{figure6}(a)
and (b) it follows that the dots are separated by the center gate
electrode. Recall that devices A and B are identical except that
for device B the center gate (instead of gate~2) is used to
control dot~2. The dashed lines in Fig.~\ref{figure6}(a) and (b)
connect triple points corresponding to a constant charge
$Q_{1(2)}=const.$, residing on dot~1(2). A non-zero slope of these
lines indicates a cross capacitance, i.e. the gate controlling one
of the two dots also affects the chemical potential of the other.
A non-zero slope is observed for the $Q_1=const.$-lines in (b).
Hence, the center~gate affects both dot~1 and dot~2. On the other
hand, this is not the case for gate~1 or gate~2 in
Fig.~\ref{figure6}(a) and (b). Such behavior can be explained
assuming that the two dots are separated by the center~gate,
screening the cross-action of gates~1~and~2. The center~gate,
however, located in between the dots and creating a barrier, is
not screened and thus acts on the two dots. If the center gate is
capable of creating a tunnel barrier inside the nanotube, so will
be gate~1 and gate~2 as well. Also recall that the voltages
applied to the top-gates are all within the same range,
$V_{Top-gate}\approx0$~V. Consequently dot~1 is located between
gate~1 and center~gate, whereas dot~2 extends from the center~gate
to gate~2. The scenario suggested implies that the part of the
SWNT between gate~1(2) and source (drain) electrode has an
effectively energy-independent transmission. In fact this
assumption is quite reasonable, taking into account the high
quality of Pd-nanotube electrical contacts~\cite{daiPd}. Very
transparent contacts ($\Gamma\approx\delta E$) lead to a constant,
or at least only slightly modulated transmission. Transport
through our device will be dominated by the bottleneck in
transmission - the gate-defined double quantum dot.

%%%%%%%%%%%%%%%%%%%%%%%%%%%%%%%%%%%%%%%%%%%%%%%%%%%%%%%%%%%%%%%%%%%%%%%%%%%%%%%%%
%%%%%%%%%%%%%% Conclusions %%%%%%%%%%%%%%%%%%%%%%%%%%%%%%%%%%%%%%%%%%%%%%%%%%%%%%
%%%%%%%%%%%%%%%%%%%%%%%%%%%%%%%%%%%%%%%%%%%%%%%%%%%%%%%%%%%%%%%%%%%%%%%%%%%%%%%%%

\section{Conclusions}

In this article we have presented a reliable approach to define
and control double quantum dots in SWNTs by using locally acting
top-gate electrodes. That the double quantum dots are not
intrinsic to the carbon nanotubes is confirmed by the presented
measurements of honeycomb patterns for three different devices.
Furthermore, using an electrostatic model, we have been able to
characterize the double-dot system quantitatively by extracting
its capacitances. Despite these encouraging results, further
research is necessary. Challenges to master include the
gate-control of the quantum-mechanical tunnel coupling of the two
quantum dots and the access to regimes of only a few charge
carriers per dot. Carbon nanotubes may, due to their unique
properties and their experimental ease, then play an important
role in future information technology.

%%%%%%%%%%%%%%%%%%%%%%%%%%%%%%%%%%%%%%%%%%%%%%%%%%%%%%%%%%%%%%%%%%%%%%%%%%%%%%%%%
%%%%%%%%%%% Acknowledgements %%%%%%%%%%%%%%%%%%%%%%%%%%%%%%%%%%%%%%%%%%%%%%%%%%%%
%%%%%%%%%%%%%%%%%%%%%%%%%%%%%%%%%%%%%%%%%%%%%%%%%%%%%%%%%%%%%%%%%%%%%%%%%%%%%%%%%

\section*{Acknowledgements}
For theoretical support we are gratefully indebted to W.A. Coish
and D. Loss. We acknowledge experimental contributions by J.
Furer, C. Hoffmann, and J. Gobrecht for oxidized Si substrates.
Financial support from the Swiss NFS, the NCCR on Nanoscience, and
the `C. und H. Dreyfus Stipendium'~(MRG) is greatly appreciated.

%%%%%%%%%%%%%%%%%%%%%%%%%%%%%%%%%%%%%%%%%%%%%%%%%%%%%%%%%%%%%%%%%%%%%%%%%%%%%%%%%%%
%%%%%%%%%%%% References %%%%%%%%%%%%%%%%%%%%%%%%%%%%%%%%%%%%%%%%%%%%%%%%%%%%%%%
%%%%%%%%%%%%%%%%%%%%%%%%%%%%%%%%%%%%%%%%%%%%%%%%%%%%%%%%%%%%%%%%%%%%%%%%%%%%%%%%%%

\section*{References}

\bibliographystyle{apsrev}

\end{document}